\documentclass[twocolumn,aps,showpacs,groupedaddress,floatfix,prl]{revtex4-1}
\usepackage{amsmath}
\usepackage{graphicx}
\usepackage{dcolumn}
\usepackage{bm}
\usepackage{verbatim}

\begin{document}
\setlength{\parskip}{0mm}
\title{Universal Scaling of Pair-Excess Entropy and Diffusion in Strongly Coupled Liquids}
\author{Ashwin Joy}
\email{ashwin@iitm.ac.in}
\affiliation{Department of Physics, Indian Institute of Technology Madras, Chennai - 600036}
\date{\today}
\begin{abstract}
Understanding diffusion in liquids from properties of static structure is a long standing problem in
condensed matter theory. Here we report an atomistic study of excess entropy and diffusion
coefficient in a strongly coupled Yukawa liquid. We observe that the pair excess entropy $s_2$ scales
with temperature as $-3.285 \;(T_m / T)^{0.665}$ and contributes to about $90\%$ of the total
excess entropy close to the freezing transition $T_m$. We further report that at low temperatures
where the diffusive transport is mediated by cage relaxation, the diffusion coefficient
when expressed in natural units of the Enskog collision frequency and the effective hard sphere
diameter, obeys the scaling law $0.04\; e^{s_2}$ and deviates from it at high enough temperatures 
where cages cannot form. The scaling laws reported here may also apply to strongly 
coupled dusty plasmas and charged colloids.
\end{abstract}

 \keywords{} \maketitle 
A unifying description of atomic diffusion in condensed matter has remained elusive so far
\cite{jakse2016excess, Bagchi-2015, Samanta-2001, rosenfeld1977relation}. For dilute gases the
Chapman-Enskog solution of the Boltzmann equation applies well as the collisions in this limit, to a
good approximation are assumed to be binary in nature \cite{chapman1970mathematical,
marrero1972gaseous}. Higher densities can be addressed after a generalization of the Enskog theory
\cite{COHEN1993229} or by using an effective Boltzmann approach \cite{PhysRevE.93.043203}. 
The late 1960s saw the first
computer simulations on transport phenomena prompting the development of mode coupling theories
which were more successful in describing diffusion in moderate to dense fluids
\cite{Alder-Wainwright-VACF1, Alder-Wainwright-VACF2, Sjogren-Sjolander-MCT}.  For liquids that are
dense or at very low temperatures, structural effects dominate over kinetic effects and change the
qualitative features of atomic transport \cite{saigo:1210, PhysRevLett.94.185002,
PhysRevLett.93.155004}. These structural effects manifest themselves in the form of a cage around a
given particle formed by nearest neighbors \cite{Weeks627, ashwin2015microscopic}. As temperature
rises, the short range order of a liquid begins to fluctuate rapidly and eliminates any caging
effect that may affect diffusive transport.  Particularly at low temperatures where dynamical
excitations become collective in nature, universal scaling laws emerge in both thermodynamics and
transport phenomena \cite{donko2004thermal, ohta2000molecular}. Whether such a bridge exists that
can connect underlying structure with dynamics in strongly coupled liquids is the subject matter of
our work.
  
Rosenfeld \cite{rosenfeld2000excess} proposed a connection between diffusion co-efficient $D$ and
the total excess entropy per particle $s$ in the form of a scaling law $D n^{1/3} / (k_B T /
m)^{1/2} = A e^{B s}$ where $n$ is the number density.  It should be noted that $s$ which arises due
to structural correlations, is over and above the ideal gas value and is therefore negative.  The
prefactor $A$ and the exponential argument $B$ vary for different inter-atomic potentials and the
scaling law is able to estimate the diffusion coefficient to within $30\%$ of the actual value.  The
scaling thus acts as a corresponding states like relationship and was observed to be only
quasi-universal in nature. The dimensional reduction of $D$ is macroscopic in nature as the
parameters $n^{1/3}$ and $(k_B T / m)^{1/2}$  do not depend on the liquid structure.  Later works
\cite{dzugutov1996universal}  showed that a microscopic reduction of $D$ using structure dependent
parameters namely the Enskog collision frequency $\nu$ and effective hard sphere diameter $\sigma$
leads to a scaling law $D / (\nu \sigma^2) = 0.049 e^{s_2}$.  Here $\sigma$ is defined as the
location of the  first peak in the pair correlation function $g(r)$ and $\nu$ is given by $4
\sigma^2 g(\sigma) n \sqrt{\pi k_B T/m}$. The pair excess entropy $s_2$ is a two body approximation
of the full configuration entropy and can be readily obtained once $g(r)$ is available (see text
later). The scaling law which works for a range of soft potentials was seen to break down in liquid
Silicon and some liquid metals \cite{hoyt2000test}. To the best of our knowledge, a microscopic
study of this pair excess entropy and its connection to the diffusion coefficient in strongly
coupled liquids with long ranged interactions is lacking. Such a study will be immediately useful
for kinetically resolved experiments in dusty plasma and charged colloids where the particle
interactions to a good approximation can be considered to be the Yukawa potential.  The purpose of
this letter is to address this issue and provide a universal scaling of pair excess entropy
(directly calculated from structure) with temperature in a strongly coupled Yukawa liquid. In what
follows we also provide a scaling law that connects this pair excess entropy with liquid diffusion.   
 
 Our prototype system is a three dimensional (3D) strongly coupled Yukawa liquid which is known to
 be an excellent model for dusty plasma and colloidal suspension of charged particulates
 \cite{RevModPhys.81.1353, Fortov20051}.  The availability of kinetically resolved in experiments in
 dusty plasma \cite{PhysRevLett.92.175004} further brightens up the prospect of a direct comparison
 with our results reported here.  The particles in the liquid interact through the Yukawa potential
 $\phi (r) = Q^2 (4 \pi \epsilon_0 r)^{-1} e^{-r/\lambda_D}$ where $Q$ is the particle charge and
 $\lambda_D$ is the Debye length of the background plasma. We have performed molecular dynamics (MD)
 simulations in a canonical ensemble with periodic boundary conditions. We neglect neutral gas friction 
 as it affects the diffusion dynamics only in the limit of large dissipation
 \cite{PhysRevE.82.036403,PhysRevE.66.016404}. Distance is expressed in
 units of the Wigner-Seitz radius $a = (4 \pi n/3)^{-1/3}$, energy in units of $Q^2 / (4 \pi
 \epsilon_0 a)$, time in units of inverse nominal dust frequency $\sqrt{3 \epsilon_0 m / (Q^2 n)}$
 and entropy in units of $N k_B$.  The system can exhibit a state of strong coupling when the
 dimensionless parameter $\Gamma = Q^2 / (4 \pi \epsilon_0 a k_B T) >> 1$ leading to a remarkable
 display of collective excitations and self organization phenomena \cite{Donko-Kalman-Hartmann-2008,
 PhysRevLett.78.3113}. The two dimensionless parameters $\Gamma$ and $\kappa = a/\lambda_D$
  completely describe the thermodynamics of our system. 
 We take 10648 particles at a reduced density $n = 3/(4 \pi)$.  To speed up the
 simulations we smoothly truncate the interaction potential along with its two derivatives to zero
 at a cutoff distance $r_c$. This is done by employing a fifth-order polynomial function that is
 switched on when $r_m < r < r_c$ with $r_m$ and $r_c$ being the inner and outer cutoff distance
 respectively. We choose $r_m$ and $r_c$ subject to criteria $\phi (r_m) \approx 4.085 \times
 10^{-4}$ and $\phi (r_c) \approx 9.300 \times 10^{-7}$ thus ensuring negligible perturbation to the
 bare Yukawa potential. A Nos\'{e}-Hoover thermostat \cite{PhysRevA.31.1695} maintains constant 
 temperature in the NVT ensemble. To improve statistics we average our data over 
 30 statistically independent realizations.  

 We begin by first computing the total excess entropy $s$ by performing a thermodynamic integration
 of the equation of state along a reversible path from some reference state $\Gamma_0$ to the
 desired $\Gamma$ as mentioned below
 \begin{equation}
   s_{\Gamma \leftarrow \Gamma_0} = u(\Gamma) \Gamma - u(\Gamma_0) \Gamma_{0} - \int_{\Gamma_0}^{\Gamma} u (\Gamma')
   \text{d}\Gamma'
   \label{eq-s-TI}
 \end{equation}
 where $u$ is the potential energy per particle. Ideally one should take $\Gamma_0$ to be zero in
 order to measure entropy relative to the ideal gas state but in practice it is difficult to carry
 the above integration at such low values of $\Gamma$. We avoid this problem by taking $\Gamma_0 =
 1$ as a reference state as all peaks in $g(r)$ disappear at this temperature.  An alternate route
 to calculate excess entropy directly from the underlying structure is due to Wallace
 \cite{wallace1987role}.  His method is based on the expansion of $s$ in terms of correlation functions
 and gives excess entropy relative to the ideal gas state. We use Ref.
  \cite{baranyai1989direct} to write for $s$:
  \begin{widetext}
    \begin{align}
  s = \underbrace{- \frac{1}{2} n \int [ g(r) \text{ln}\{g(r)\} + \{1 - g(r)\}] d \textbf{\text{r}}}_{\mbox{$s_2$}} 
  \underbrace{- \frac{1}{6} n^2 \int\int [g^{(3)}(r)\text{ln}\{\delta g^{(3)}(r)\} + 3
      {g(r)}^2 - g^{(3)}(r) - 3g(r) + 1] d {\textbf{\text{r}}}^2}_{\mbox{$s_3$}} +
      \cdots
    \label{eq-s-Wallace}
  \end{align}
  \end{widetext}
  \noindent
\begin{figure}[h] \centering
  \includegraphics[scale=0.39]{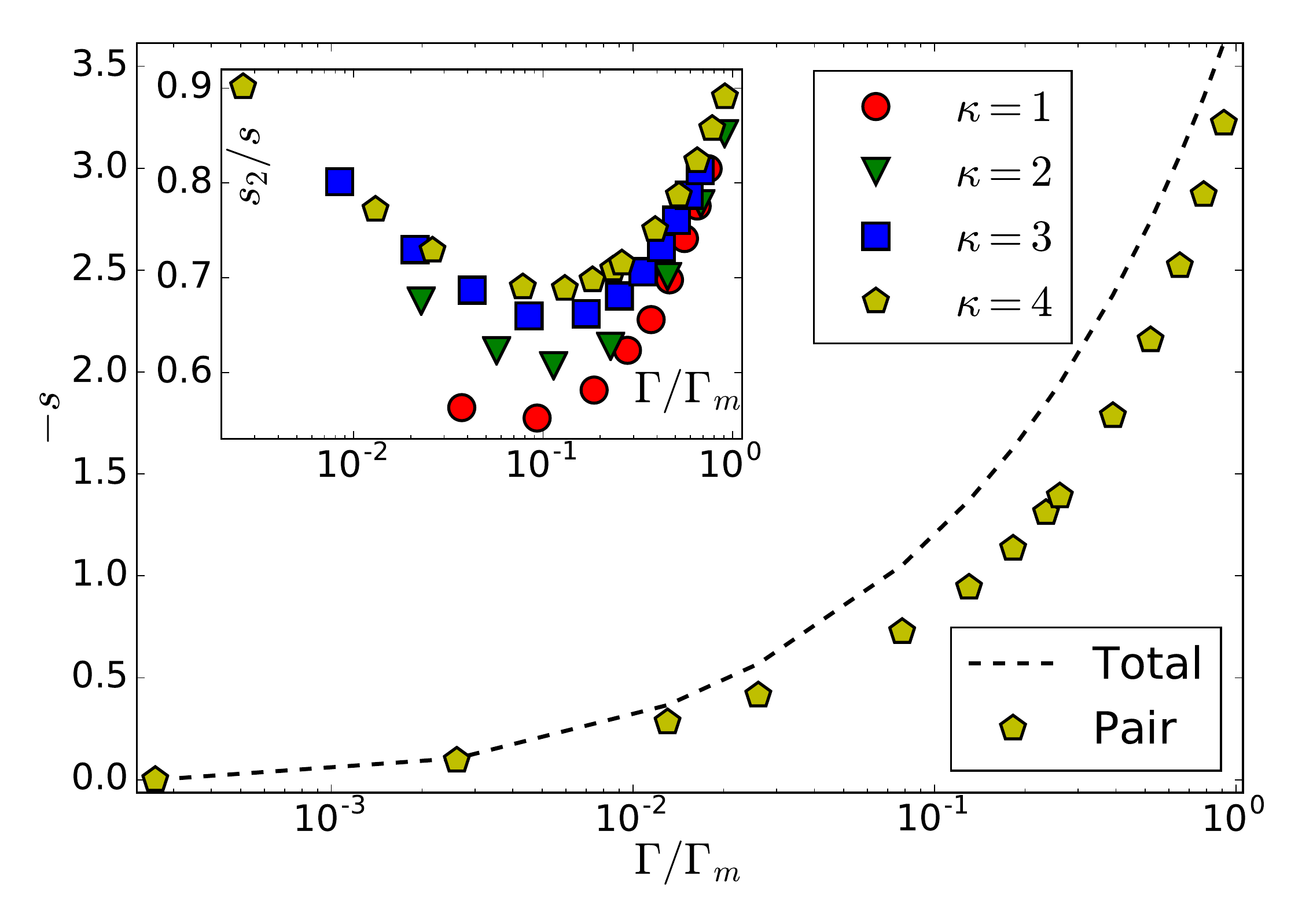} \caption{(color online). Comparison of the 
    total excess entropy [Eq. (\ref{eq-s-TI})] with the pair contribution [Eq. (\ref{eq-s-Wallace})]
    shown for the case $\kappa = 4$ and the reference state $\Gamma_0 = 1$. Inset: Fraction of
the total excess entropy as a function of $\Gamma$.} 
  \label{fig-s2-vs-s} 
\end{figure} 
\begin{figure}[h] \centering
\includegraphics[scale=0.39]{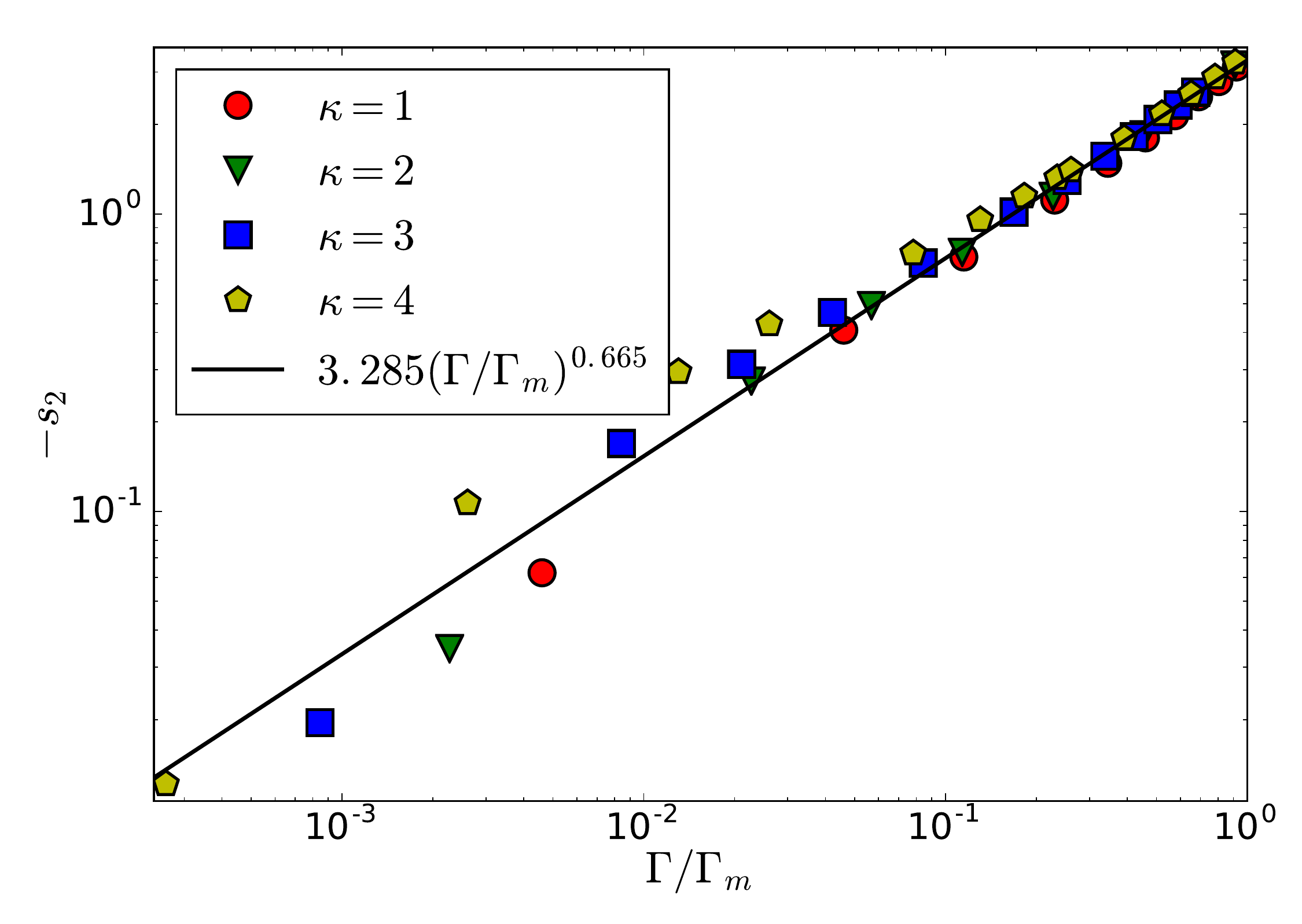} \caption{(color online). Pair excess entropy as a
  function of reduced inverse temperature $\Gamma / \Gamma_m$. Data collapse 
indicates the accuracy of our scaling law down to $\Gamma$ as low as $2\%\; \Gamma_m$.}
\label{fig-s2-scaling} 
\end{figure} 
 Here $s_2, s_3, \cdots$ are the two body, three body and other many body contributions to the total
 entropy $s$ respectively.  In Fig. \ref{fig-s2-vs-s} we plot a comparison of $s$ [from Eq.
 (\ref{eq-s-TI})] with $s_2$ keeping $\Gamma_0 = 1$ as the reference.  We find that close to freezing
 $s_2$ dominates over other many body terms and contributes to over $90 \%$ of $s$ [see Fig.
 \ref{fig-s2-vs-s}: Inset]. Note the existence of a minimum around $\Gamma = 10\%\Gamma_m$ and
 highest contribution at $\Gamma$ extrema.  
\begin{figure}
\includegraphics[scale=0.39]{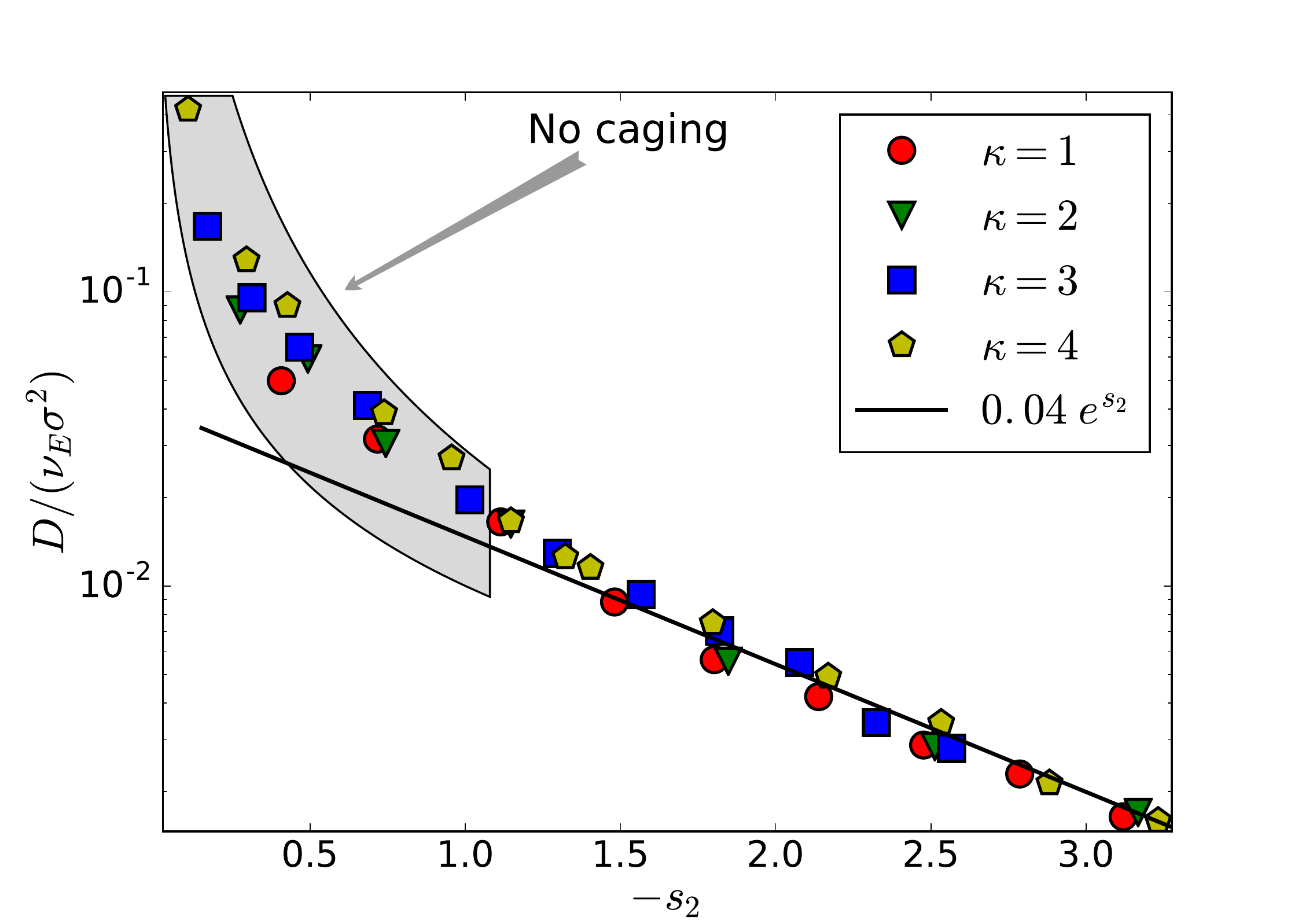} \caption{(color online). Reduced diffusion
  coefficient vs pair excess entropy. Except when caging is absent (shaded region) we are able to
  collapse our data on to the simple scaling law shown by a solid line.  Data inside the shaded region are taken
  at temperatures where the velocity auto-correlation does not cross zero [see text and Fig.
  \ref{fig-vacf}].} \label{fig-D-scaling} 
\end{figure} 
Next we show the scaling of this pair excess entropy with the reduced temperature $\Gamma /
\Gamma_m$ in Fig. \ref{fig-s2-scaling}. It should be noted that $\Gamma_m$ varies by more than an
order of magnitude from $217.4$ to $3837$ as $\kappa$ goes from 1 to 4 \cite{PhysRevE.56.4671}
and hence we have sufficient reasons to believe the scaling law $s_2 = - 3.285
(\Gamma/\Gamma_m)^{0.665}$ reported here is universal in nature. The value of $s_2$ at melting 
$\approx -3.285$ for all $\kappa$. Our results may thus prove to be very useful in predicting structural
entropy from trajectory snapshots which are easily available in kinetically resolved dusty plasma experiments.
In what follows we will provide a link to connect this structural information to the dynamics that
governs transport.

Any theory that unifies dynamical properties such as diffusion with the underlying structure will
need to provide ways in which the local arrangement of nearest neighbors (or cage) around a
particle may affect its long time dynamics. Recently we showed that the lifetime of this cage
decides the relaxation of shear stress in the liquid state \cite{ashwin2015microscopic}. 
Thus it is natural to expect that cage relaxation will also be necessary 
to produce local density fluctuations necessary for diffusive transport especially at
low temperatures where the liquid exhibits strong caging behavior. Since a cage is typically
 formed by the nearest neighbors [particles within the first peak of $g(r)$],
the momentum and energy transfer processes between particles can be expected to be short
ranged just like binary collisions in a gas of hard spheres. The relevant time scale of these
collisions can then be given by the inverse of Enskog collision frequency $\nu_E$ and the relevant
length scale can be realized as the effective hard sphere diameter $\sigma$ which is just the
location of the first peak in $g(r)$. Within the classical Enskog theory we have 
\begin{equation}
  \nu_E = 4 \sigma^2 g(\sigma) n \sqrt{\pi k_B T/m}
  \label{eq-nu_E}
\end{equation}
\begin{figure}[h] \centering
  \includegraphics[scale=0.39]{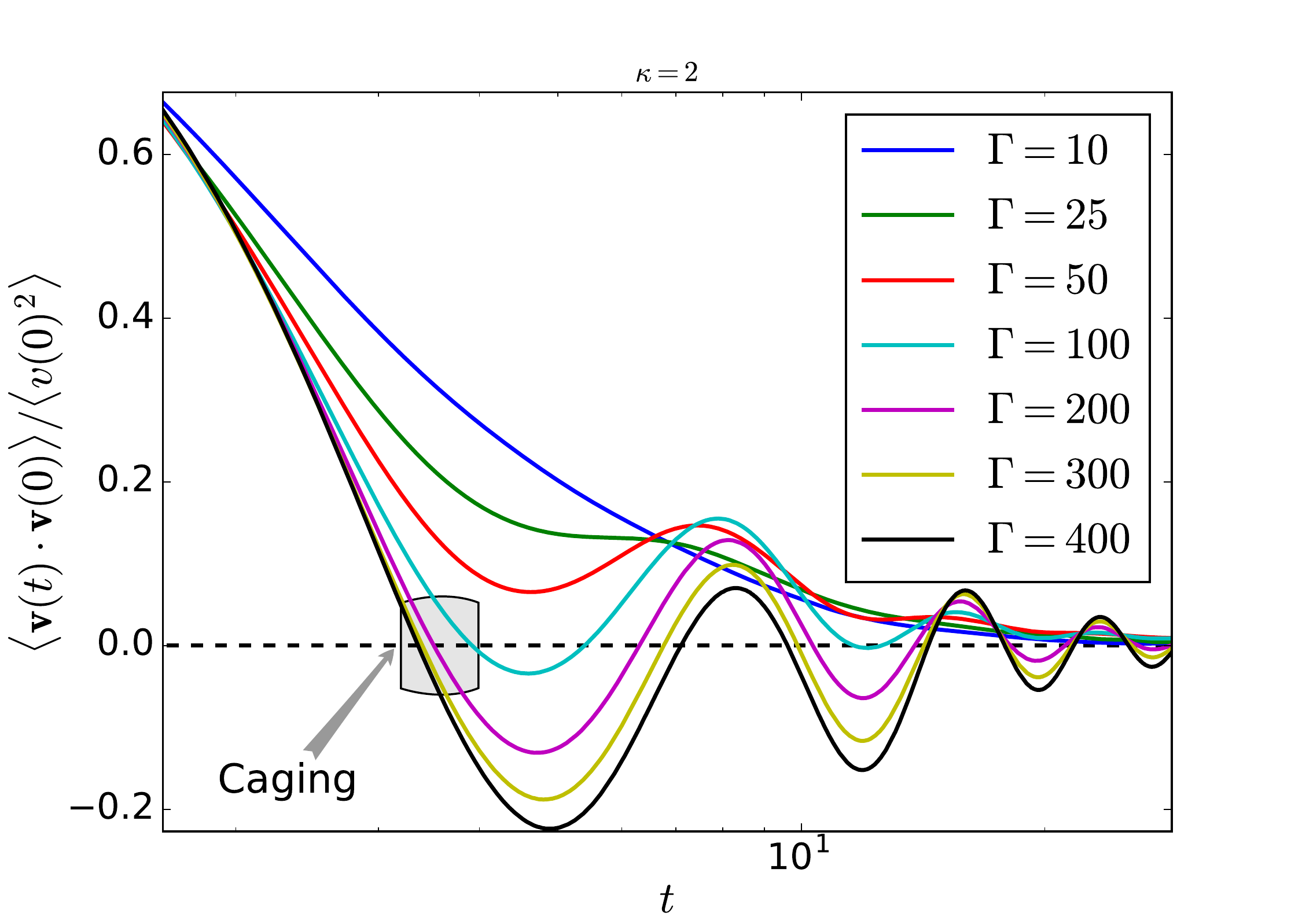} \caption{(color online). Normalized velocity
    auto-correlation data for the case $\kappa = 2$. Zero crossing of the function indicates caging
    behavior. For $\Gamma$ values 10, 25 and 50 where caging is absent we see deviations from the
    scaling law as shown in the shaded region of Fig.  \ref{fig-D-scaling}.} \label{fig-vacf}
\end{figure} 
We now turn our attention to Fig. \ref{fig-D-scaling} where we plot the diffusion coefficient in
units of $\nu_E \sigma^2$ as a function of pair excess entropy and find that a scaling law $D /
(\nu_E \sigma^2) = 0.04 e^{s_2}$ holds in the region $-s_2 > 1$. However when $-s_2 < 1$, a
deviation from this scaling law is observed which is shown by the shaded region. Our prefactor of
$0.04$ which is about $18\%$ smaller than the one used in Ref. \cite{dzugutov1996universal} is not
universal.  Note that lower values of $-s_2$ imply higher temperatures. The points that deviate from
the scaling law are taken at temperatures high enough to set rapid fluctuations in the short range
order of the liquid thus marginalizing any role that the nearest neighbor dynamics can play in
particle diffusion. Our data for velocity auto-correlation (VAC) for the case $\kappa = 2$ shown
in Fig. \ref{fig-vacf} confirms this. Data for other values $\kappa$ are similar (not shown here).
When the liquid exhibits caging, indicated by zero-crossing of VAC, we find that the corresponding
data for diffusion obeys the scaling law shown in Fig. \ref{fig-D-scaling}. This is expected because
at these temperatures cage relaxation is necessary for the onset of diffusive transport. At
temperatures where the VAC does not cross zero, there is no caging and $D$ does not obey the scaling
law. These data points are shown in the shaded region of Fig. \ref{fig-D-scaling}.

\begin{figure}[h] \centering \includegraphics[scale=0.39]{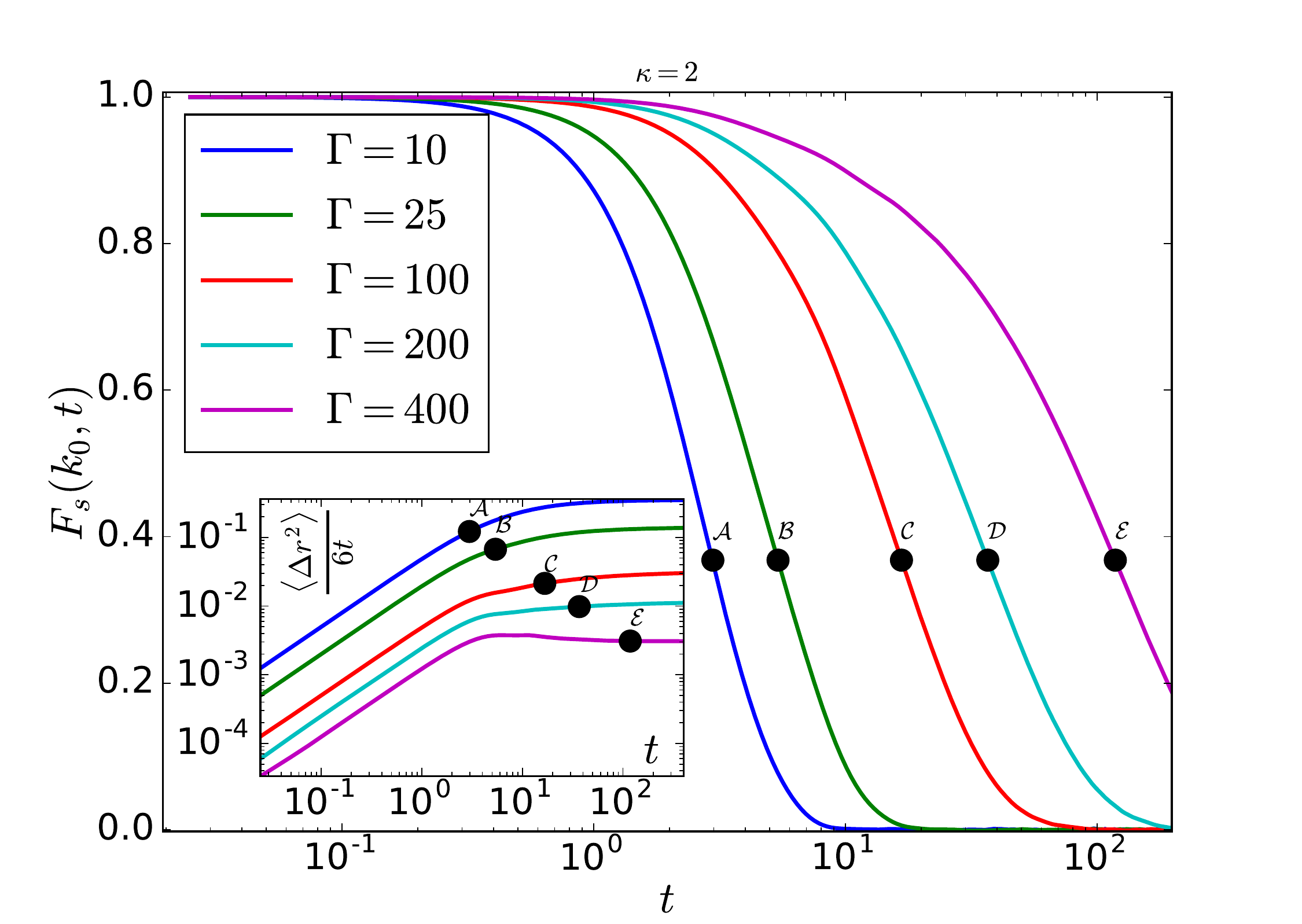}
 \caption{(color online). Self intermediate scattering function $F_s(k_0, t)$ at various $\Gamma$
 for the case $\kappa = 2$. We take $k_0 = 2\pi/\sigma$ with $\sigma$ being the location of the 
 first peak in $g(r)$. Points $\mathcal{A, B, C, D}$ and $\mathcal{E}$ mark the e-folding times of $F_s(k_0,
 t)$. Inset: Diffusion coefficient shown for the same $\Gamma$.}
 \label{fig-sisf-msd} 
\end{figure}

To further strengthen our argument, we provide in Fig. \ref{fig-sisf-msd} plots of
self-intermediate scattering function $F_s(k_0, t)$ at various $\Gamma$ for the case $\kappa =
2$. The value of $k_0$ is taken to be $2\pi/\sigma$. The points $\mathcal{A, B, C, D}$ and
$\mathcal{E}$ denote the locations of the e-folding times ($t_\alpha$) in the $F_s(k_0, t)$ data. As
$t_\alpha$ tells us how long one must typically wait for the cages to break or rearrange, it is a
good measure of the structural relaxation time. Thus the diffusive regime if mediated by cage
relaxation must occur at $t \gtrsim t_\alpha$.  Indeed we observe this in the inset where we show
the corresponding plots of the diffusion coefficient with the same points marking the onset of cage
breaking. We notice that for the cases $\Gamma = 100, 200$ and $400$ where cages do form (see Fig.
\ref{fig-vacf}) the corresponding points $\mathcal{C, D}$ and $\mathcal{E}$ mark the onset of
diffusive regime (i.e $D \approx$ constant). Diffusive transport at these values of $\Gamma$ is thus
mediated by cage relaxation and the diffusion data obeys our scaling law shown in Fig.
\ref{fig-D-scaling}. At $\Gamma = 10$ and $25$, the liquid does not exhibit any caging behavior and
hence diffusion cannot be mediated by local structural relaxation- also confirmed from the location
of points $\mathcal{A}$ and $\mathcal{B}$ in the inset of
Fig. \ref{fig-sisf-msd}. Diffusion data at these $\Gamma$ thus deviate from the scaling
$D/(\nu\sigma^2) = 0.04 e^{s_2}$.  

We can conclude that the pair excess entropy $s_2$ in liquid dusty plasma scales as $- 3.285 (T_m /
T)^{0.665}$ for all values of screening parameter with $-3.285$ being the value of $s_2$ at melting
for all screening lengths.  Especially close to the melting point $T_m$, we
find that $s_2$ which arises from pair correlation contributes to almost $90\%$ of the total
entropy. Our scaling is universal as the melting points vary by more than an order of magnitude over
the entire range of screening parameters used in this work.  We further report that at low
temperatures where the liquid exhibits caging, the diffusion coefficient when expressed in the
natural units of the Enskog collision frequency and the effective hard core diameter scales as
$0.04\; e^{s_2}$. At higher temperatures when caging is absent, deviation from the scaling law is seen. The
prefactor in our scaling law for diffusion is different from the one in past works 
\cite{dzugutov1996universal} implying that the prefactor itself in not universal. The scaling laws reported here may
prove to be valuable in directly computing excess entropy and diffusion coefficient in kinetically
resolved liquid dusty plasma experiments where particle snapshots are easily obtained.
\begin{acknowledgments} The author wishes to thank Abhijit Sen and Rajaraman Ganesh for
  discussions. All simulations were done on the VIRGO super cluster of IIT Madras.
\end{acknowledgments}

\bibliography{ashwin}

\end{document}